 \documentclass[aps, twocolumn, showpacs, amsmath]{revtex4-2}

\usepackage{bm}
\usepackage{graphicx}
\usepackage{color}
\usepackage{hyperref}
\usepackage{amsthm}
\usepackage{amssymb}
\usepackage[sort&compress]{natbib}

\begin{document}

\def\be{\begin{equation}}
\def\ee{\end{equation}}

\def\bc{\begin{center}}
\def\ec{\end{center}}
\def\bea{\begin{eqnarray}}
\def\eea{\end{eqnarray}}
\newcommand{\avg}[1]{\langle{#1}\rangle}
\newcommand{\Avg}[1]{\left\langle{#1}\right\rangle}

\def\ie{\textit{i.e.}}
\def\etal{\textit{et al.}}
\def\m{\vec{m}}
\def\G{\mathcal{G}}

\newcommand{\davide}[1]{{\bf\color{blue}#1}}
\newcommand{\gin}[1]{{\bf\color{green}#1}}

\title{First-passage and extreme value statistics for overdamped Brownian motion in a linear potential}
\author{Feng Huang$^1$}
\author{Hanshuang Chen$^2$}\email{chenhshf@ahu.edu.cn}
\affiliation{$^1$School of Mathematics and Physics \& Key Laboratory of Architectural Acoustic Environment of Anhui Higher Education Institutes, Anhui Jianzhu University, Hefei 230601, China \\ $^2$School of Physics and Optoelectronic Engineering, Anhui University, Hefei 230601, China}
\begin{abstract}
We investigate the first-passage properties and extreme-value statistics of an overdamped Brownian particle confined by an external linear potential $V(x)=\mu |x-x_0|$, where $\mu>0$ is the strength of the potential and $x_0>0$ is the position of the lowest point of the potential, which  coincides with the starting position of the particle. The Brownian motion terminates whenever the particle passes through the origin at a random time $t_f$. Our study reveals that the mean first-passage time $\langle t_f \rangle$ exhibits a nonmonotonic behavior with respect to $\mu$, with a unique minimum occurring at an optimal value of $\mu \simeq 1.24468D/x_0$, where $D$ is the diffusion constant of the Brownian particle.
Moreover, we examine the distribution $P(M|x_0)$ of the maximum displacement $M$ during the first-passage process, as well as the statistics of the time $t_m$ at which $M$ is reached. Intriguingly, there exists another optimal $\mu \simeq 1.24011 D/x_0$ that minimizes the mean time $\langle t_m \rangle$. All our analytical findings are corroborated through numerical simulations.
\end{abstract}

\maketitle



\section{Introduction}\label{sec1}
Although extreme events are infrequent, their impacts can be significant, leading to destructive disasters such as earthquakes, floods, and extreme weather, causing turbulence in financial markets due to stock collapses, and triggering outbreaks of epidemics \cite{fisher1928limiting,gumbel1958statistics,leadbetter2012extremes,bouchaud1997universality,davison2015statistics,albeverio2006extreme,dumonteil2013spatial}. Extreme value statistics (EVS) is a crucial tool for studying the statistical characteristics of these rare events.  For independent and identically distributed random variables, it is well-established that the extreme-value distribution falls into three renowned universality classes: Gumbel, Fr\'echet, and Weibull\cite{gnedenko1943distribution}, depending on the tails of the distribution
of random variables. EVS has found widespread applications in diverse fields such as crystal growth \cite{kim1997dynamics}, molecular-beam epitaxy \cite{barabasi1995fractal},  the growth of interfaces \cite{majumdar2004exact,raychaudhuri2001maximal},  random matrices \cite{dean2006large,majumdar2014top}, epidemic spreads \cite{dumonteil2013spatial}, search algorithms \cite{krapivsky2000traveling,majumdar2002extreme,majumdar2003traveling,majumdar2009large}, and stochastic transport models \cite{jacobsen2010exact,guillet2020extreme}.

While the EVS of uncorrelated random variables have been thoroughly established, practical systems often involve correlated random variables \cite{majumdar2020extreme,majumdar2024statistics}. Brownian motion serves as a paradigmatic model for strongly correlated systems, and the EVS of Brownian motion, both in unbounded and bounded scenarios, has garnered increasing attention in recent years \cite{majumdar2010universal,schehr2014exact,lacroix2020universal,PhysRevLett.111.240601,PhysRevLett.117.080601,PhysRevLett.129.094101}. For a free Brownian particle initiating its journey from the origin over a time interval $ \left[ {0,t} \right]$, the distribution of its maximum value $M$ reached along its trajectory conforms to a one-sided Gaussian distribution \cite{schehr2010extreme}
\begin{eqnarray}\label{eq0.1}
P( M|t )=\frac{\Theta ( M )}{\sqrt{\pi Dt}}e^{ -\frac{M^2 }{4Dt}} ,
\end{eqnarray} 
where $D$ is the diffusion coefficient, and $\Theta \left( z \right)$ is the Heaviside step function such that $\Theta \left( z \right)=1$ if $z>0$ and $\Theta \left( z \right)=0$ otherwise. The distribution of the time ${{t}_{m}}$ at which the trajectory reaches its maximum value is given by the famous arcsine law \cite{levy1940certains,feller1971introduction,majumdar2007brownian},
\begin{eqnarray}\label{eq0.2}
P(t_{m}|t)=\frac{1}{\pi \sqrt{ t_m (t-t_m)  }}  .
\end{eqnarray} 
This finding is somewhat counterintuitive, as the probability density peaks at values of $t_m$ deviating from $t/2$, with higher likelihoods observed near the extrema $t_m=0$ and $t_m=t$, rather than at the mean value of $t_m$, $t_m=t/2$. Subsequent research has delved into the EVS of numerous extensions of Brownian motion, encompassing constrained Brownian motions \cite{majumdar2008time}, run-and-tumble particle \cite{majumdar2010random,singh2022mean}, random acceleration process \cite{majumdar2010time} , fractional Brownian motion \cite{wiese2011perturbation,delorme2016extreme,sadhu2018generalized}, anomalous walker \cite{majumdar2010hitting}, stochastic resetting systems \cite{singh2021extremal,PhysRevE.103.022135},  and Brownian motion in confined geometries \cite{chupeau2015convex,chupeau2015mean,huang2024extremal}. 

While much of the research on the EVS of Brownian motion focused on scenarios with a fixed time interval, the study of EVS for first-passage trajectories-where the observation time is a random variable rather than fixed-has also attracted growing attention in recent years. The first-passage problem is fundamental to numerous critical phenomena, including chemical reactions, animal foraging, and neural firing \cite{redner2001guide}. The EVS of stochastic processes up to a first-passage time is pertinent to various practical issues, such as determining the maximum queue length and the corresponding time before the queue empties in queue theory \cite{kearney2004random}, finding the optimal time window to sell a stock before its price drops to a threshold \cite{majumdar2008optimal}, and assessing the maximal excursions of tracer proteins before they bind to a site \cite{RevModPhys.83.81,PhysRevX.7.011019}. 
Notably, the problem of finding the maximum for first-passage trajectories is formally equivalent to computing the splitting probability, a central concept in first-passage theory that quantifies the likelihood of exiting an interval through a specific boundary. One can refer to the book of Redner \cite{redner2001guide} for more details, and to a recent work by Klinger \textit{et al.}  \cite{PhysRevLett.129.140603}, witnessing the ongoing interest in such a quantity.
This connection between splitting probability and first-passage maxima has been explored in several works \cite{randon2007distribution,guo2023extremal,guo2024extremal}. For standard Brownian motion, the distribution of first-passage times is a well-established result. In the long-time limit, this distribution decays following a power law with an exponent of $-3/2$. The EVS for an ensemble of first-passage trajectories of Brownian motions has been explored \cite{randon2007distribution,kearney2005area}. Specifically, the distribution of the maximum displacement $M$ exhibits a power-law behavior. Moreover, the time $t_m$
at which this maximum occurs displays a power-law decay with an exponent of $-3/2$ for large $t_m$ and $-1/2$ for small $t_m$. 
Recently, the EVS of first-passage trajectories has been extended to $N$ independent Brownian motions \cite{krapivsky2010maximum}, run-and-tumble particle \cite{singh2022extremal}, and resetting Brownian motion \cite{guo2023extremal,guo2024extremal,huang2024extreme}. By leveraging the EVS of excursions, the statistics of the maximum for a particular discrete model-which, in the continuous limit, is equivalent to Brownian motion in a logarithmic central potential-were derived \cite{artuso2022extreme}.
A related observable of EVS, i.e., the number of distinct sites visited by the walker before hitting a target, and the joint distribution with the first-passage time to the target were obtained analytically for random walks on one-dimensional lattices \cite{PhysRevE.103.032107,PhysRevE.105.034116}. The statistics of the functionals along the first-passage trajectories has recently  studied in various systems including Brownian motion \cite{kearney2005area,kearney2007first,majumdar2020statistics}, Brownian motion with stochastic resetting \cite{JPA2022.55.234001,DubeyJPA2023,kearney2021statistics,PhysRevE.108.044151}, and non-Markovian processes \cite{PhysRevE.107.064122,PhysRevE.105.064137,hartmann2023first}.

In this work, our objective is to investigate the EVS of first-passage trajectories of Brownian motion within a confined potential. The confinement imposed by the potential can steer the system towards an equilibrium state, thereby weakening the correlation between successive positions of the Brownian particle. A quintessential model in this context is the Ornstein-Uhlenbeck (OU) process, which represents Brownian motion subject to a harmonic potential. Prior studies have examined the EVS of the OU process for a fixed duration $t$. Notably, in the short-time limit, the mean extreme position behaves analogously to an unconfined Brownian motion, whereas in the long-time limit, the mean extreme position 
exhibits a slower growth rate of $\sqrt {\ln{t}} $ \cite{majumdar2020extreme}, rather than $\sqrt{t}$ for free Brownian motion. Here, we focus on a Brownian particle subjected to a linear external potential $V(x) = \mu |x-x_0|$, and analyze the first-passage properties of the EVS of its trajectory prior to crossing the origin, where $x_0$ represents the initial position of the particle. Our primary concern is to ascertain how the potential influences the time of first arrival at the origin $t_f$, the maximum position $M$ reached by the particle, and the time $t_m$ at which $M$ is attained. Intriguingly, our findings reveal the existence of optimal potential strengths that minimize both $t_f$ and $t_m$.

This paper is organized as follows. In Section \ref{sec2} the Brownian motion model in the linear potential is defined. The propagator of the confined Brownian motion and the mean first-passage time are deduced in Section \ref{sec3}. The distribution of the maximum position $M$ is obtained in Section \ref{sec4}. In Section \ref{sec5} we obtain the statistics of the extreme time $t_m$. Finally, the main conclusions are shown in Section \ref{sec6}.

\section{Model}\label{sec2}
\begin{figure}
	\centerline{\includegraphics*[width=1.0\columnwidth]{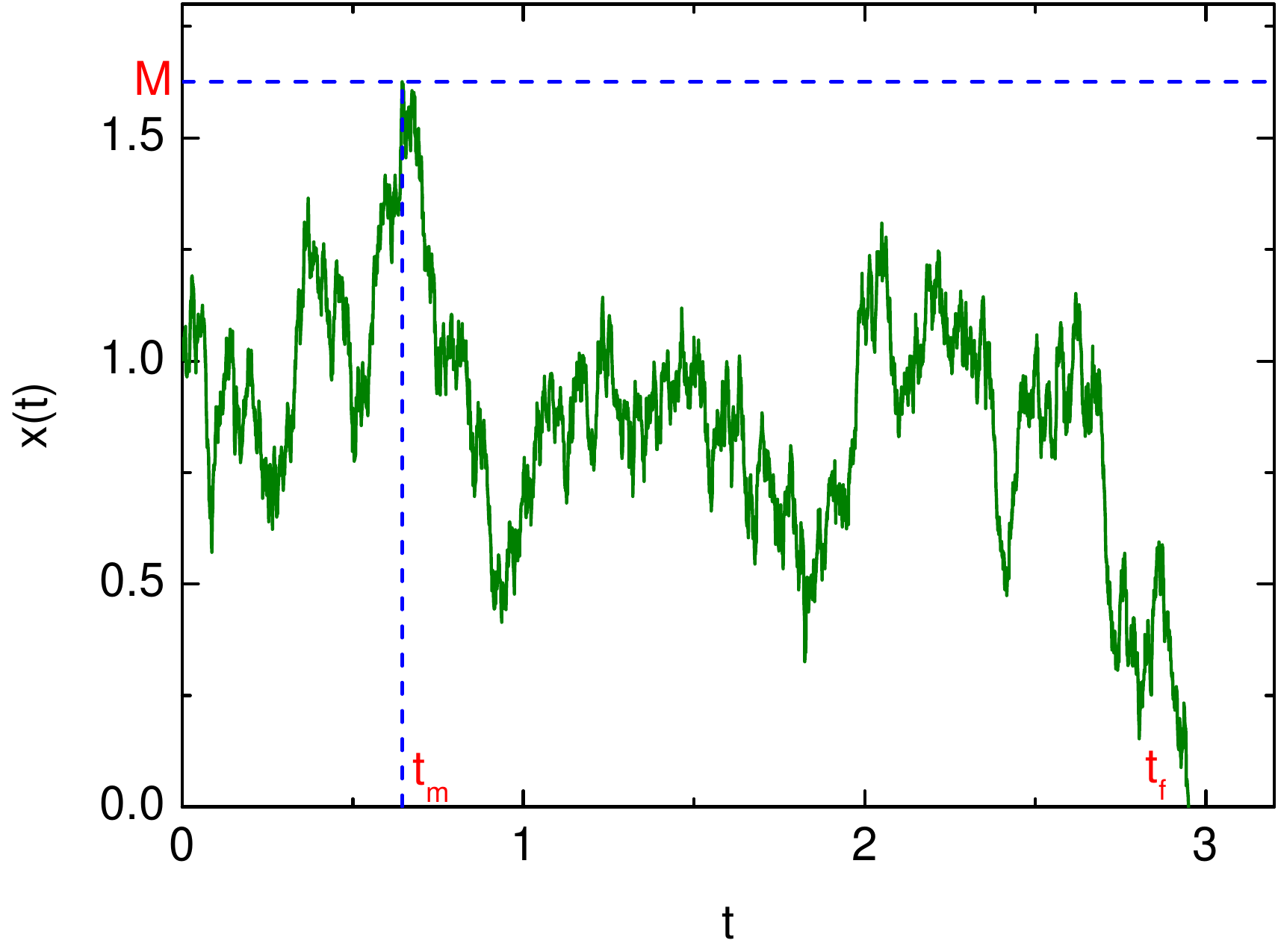}}
	\caption{The displacement $x(t)$ of overdamped Brownian
		motion in a confined linear potential $V(x)=\mu |x-x_0|$. The particle starts from $x_0>0$ and reaches its maximum displacement $M$ at time $t_m$ before the first-passage time $t_f$ through the origin. \label{fig0}}
\end{figure}

We consider the dynamics of an overdamped Brownian particle trapped by an external potential $V(x)=\mu |x-x_c|$, as described by the Langevin equation
\begin{eqnarray}\label{eq1.0}
\dot x (t)= -\mu {\rm{sgn}} (x-x_c) +\sqrt{2D}\xi (t),
\end{eqnarray} 
where $\mu>0$ is the strength of the confined potential, ${\rm{sgn}}(x)$ is the sign function such that ${\rm{sgn}}(x)>0$ if $x>0$ and ${\rm{sgn}}(x)<0$ if $x<0$,  $\xi(t)$ is the Gaussian white noise with  $\langle {\xi (t)} \rangle  = 0$ and $\langle {\xi (t)}{\xi (t')} \rangle  = \delta ( {t - t'} )$, and $D$ is the diffusion constant. The Brownian particle is assumed to be started from $x_0>0$, and terminates whenever the particle passes through the origin at a random time $t_f$. During the first-passage process, the position of the particle $x(t)$ reaches its maximum $M$ at the time $t_m$, as shown in Fig.\ref{fig0}. We are interested in the
distributions of $M$, $t_m$ and $t_f$ under the influence of the potential. 

\section{propagator and the mean first-passage time}\label{sec3}
Let us denote the propagator by $G(x,t|x_0)$, which is the probability density of finding particle at the position $x$ at time $t$ under two absorbing boundaries at $x=0$ and $x=M$, providing that the particle started from $x_0$ at $t=0$. The forward equation for $G(x,t|x_0)$ reads
\begin{eqnarray}\label{eq2.1}
	\frac{{\partial G( {x,t|{x_0}} )}}{{\partial t}} &=& D\frac{{{\partial ^2}G( {x,t|{x_0}} )}}{{\partial {x^2}}} \nonumber \\ &+& \mu \frac{\partial }{{\partial x}}\left[ {{\rm{sgn}} ( {x - {x_0}} )G( {x,t|{x_0}} )} \right],
\end{eqnarray} 
subject to boundary conditions 
\begin{eqnarray}\label{eq2.2}
	G( {0,t|{x_0}} ) = G( {M,t|{x_0}} ) = 0,
\end{eqnarray} 
and initial condition $G\left( {x,0|{x_0}} \right) = \delta \left( {x - {x_0}} \right)$, where we have assumed that the position of the potential well and the initial position of the particle are the same, $x_c=x_0$.

Performing the Laplace transformation,  $\tilde{G}(x,s|x_0)=\int_{0}^{\infty} G(x,t|x_0) e^{-st}dt$, Eq.(\ref{eq2.1}) now becomes
\begin{eqnarray}\label{eq2.4}
	s\tilde G( {x,s|{x_0}} ) - \delta ( {x - {x_0}} ) &=& D\frac{{{d^2}\tilde G( {x,s|{x_0}} )}}{{d{x^2}}} \nonumber \\& +& \mu \frac{d}{{dx}}\left[ {{\rm{sgn}} ( {x - {x_0}} )\tilde G( {x,s|{x_0}} )} \right].\nonumber \\
\end{eqnarray} 
Using boundary conditions in Eq.(\ref{eq2.2}), Eq.(\ref{eq2.4}) can be solved for $x<x_0$ and $x>x_0$, separately, 
\begin{widetext}
\begin{eqnarray}\label{eq2.5}
\tilde G (x,s|x_0) =\left\{ \begin{array}{lll}{c_1}\sinh \left( {\omega _s}x\right)\left[ {\sinh \left( \hat \mu x \right) + \cosh \left( \hat \mu x \right)} \right],    & x<x_0,  \\
{c_2}\sinh \left[ {\omega _s}(M-x) \right]\left[ {\sinh \left( {\omega _s}M +\hat \mu x \right) - \cosh \left( {\omega _s}M +\hat \mu x \right)} \right],    & x>x_0,  \\ 
\end{array}  \right. 
\end{eqnarray}
where
\begin{eqnarray}\label{eq2.6}
	{\omega _s} = \frac{\sqrt {4Ds + {\mu ^2}}}{2D},
\end{eqnarray}
and
\begin{eqnarray}\label{eq2.6.1}
\hat \mu=\frac{\mu}{2D}.
\end{eqnarray}
To obtain two unknown coefficient $c_1$ and $c_2$ in Eq.(\ref{eq2.5}), we seek to two continuity conditions at $x=x_0$. The first one is $\tilde G (x_0^+,s|x_0)=\tilde G (x_0^-,s|x_0)$, and the second one can be obtained by integrating over Eq.(\ref{eq2.4}) from $x_0-\epsilon$ to $x_0+\epsilon$, and then letting $\epsilon \to 0$,  leading to
\begin{eqnarray}\label{eq2.7}
	- 1 = D\left[ {\tilde G'\left( {x_0^ + ,s|{x_0}} \right) - \tilde G'\left( {x_0^ - ,s|{x_0}} \right)} \right] + \mu \left[ {\tilde G\left( {x_0^ + ,s|{x_0}} \right) + \tilde G\left( {x_0^ - ,s|{x_0}} \right)} \right].
\end{eqnarray}
From the two continuity conditions, we obtain $c_1$ and $c_2$, and then insert them into Eq.(\ref{eq2.5}), to have
\begin{eqnarray}\label{eq2.8}
	\tilde G (x,s|x_0) =\frac{{\sinh \left( {{\omega _s}{x_<}} \right)\sinh \left[ {{\omega _s}\left( {M - x_>} \right)} \right]e^{{\mu \left( {{x_ < } - {x_ > }} \right)}} }}{{D\left( {\hat \mu \cosh \left[ {{\omega _s}\left( {M - 2{x_0}} \right)} \right] - \hat \mu \cosh \left( {{\omega _s}M} \right) + {\omega _s}\sinh \left( {{\omega _s}M} \right)} \right)}},
\end{eqnarray}
\end{widetext}
where $x_<=\min\left\lbrace x, x_0 \right\rbrace $ and $x_>=\max\left\lbrace x, x_0 \right\rbrace $.
The probability current through the origin can be written as
\begin{eqnarray}\label{eq2.8.1}
	J\left( {t,{0}} |x_0\right) = {\left. {D\frac{{\partial G\left( {x,t|{x_0}} \right)}}{{\partial x}} } \right|_{x = 0}}.
\end{eqnarray}
and its Laplace transformation $\tilde{J}\left( {s,{x_0}} \right) = \int_{0}^{\infty} J\left( {t,{0}}|x_0 \right) e^{-st} dt$  can be obtained by using Eq.(\ref{eq2.8}), 
\begin{eqnarray}\label{eq2.8.2}
	&&\tilde{J}( {s,{0}} |x_0 ) = {\left. {D\frac{{\partial \tilde{G}( {x,s|{x_0}} )}}{{\partial x}} } \right|_{x = 0}}   \nonumber \\&=&
	\frac{{{\omega _s}{e^{-\hat \mu {x_0}}}\sinh \left[ {{\omega _s}\left( {M - 2{x_0}} \right)} \right]}}{{\hat \mu \cosh \left[ {{\omega _s}( {M - 2{x_0}} )} \right] - \hat \mu \cosh ( {{\omega _s}M} ) + {\omega _s}\sinh ( {{\omega _s}M} )}}. \nonumber \\
\end{eqnarray}
Taking the limit of $M \to \infty$ for Eq.(\ref{eq2.8.2}), we obtain the first-passage probability distribution through the origin in the Laplace domian, 
\begin{eqnarray}\label{eq2.8.31}
	\tilde F( {s,{x_0}} ) &=& \mathop {\lim }\limits_{M \to \infty} \tilde{J}\left( {s,{0}} |x_0 \right) \nonumber \\ &=&\frac{{{\omega _s}{e^{\left( {{\omega _s} - \hat \mu } \right){x_0}}}}}{{\hat \mu  - \hat \mu {e^{2{\omega _s}{x_0}}} + {\omega _s}{e^{2{\omega _s}{x_0}}}}}.
\end{eqnarray}
The moments of first-passage time through the origin starting from $x_0$ can be obtained by calculating, 
\begin{eqnarray}\label{eq2.8.3.0}
	\langle {{t_f^n}} \rangle  =  (-1)^n \mathop {\lim }\limits_{s \to 0 } \frac{{\partial^n \tilde F\left(s, x_0 \right)}}{{\partial s^n}} .
\end{eqnarray}
In particular, the mean first-passage time is given by
\begin{eqnarray}\label{eq2.8.3}
	\langle {{t_f}} \rangle  = \frac{{2D\left( {{e^{\mu {x_0}/D}} - 1} \right) - \mu {x_0}}}{{{\mu ^2}}},
\end{eqnarray}
In the limit of $\mu \to 0$, $\langle {{t_f}} \rangle$ diverges as $\langle {{t_f}} \rangle \sim x_0/\mu$. As $\mu$ increases, $\langle {{t_f}} \rangle$ first decreases and then increases, and thus there is an optimal $\mu=\mu_{{\rm{opt}}}$ such that $\langle {{t_f}} \rangle$ is a minimum, where $\mu_{{\rm{opt}}}$ can be obtained by numerically solving a transcendental equation ${e^{ - y_{{\rm{opt}}}}} = \frac{{4 - 2y_{{\rm{opt}}}}}{{y_{{\rm{opt}}} + 4}}$ ($y_{{\rm{opt}}}=\mu_{{\rm{opt}}} x_0/D>0$), yielding
\begin{eqnarray}\label{eq2.8.4}
\mu_{{\rm{opt}}} \simeq 1.24468 \frac{D}{x_0}.
\end{eqnarray} 
We should note that results in Eqs.(\ref{eq2.8.3})  and (\ref{eq2.8.4}) were already obtained in a previous work by Mercado-V{\'a}squez \textit{et al.} \cite{mercado2020intermittent}, where the authors considered the first-passage problem of Brownian motion in a switching $V$-shaped potential as a physical realization of stochastic resetting \cite{evans2020stochastic}.

Similarly, we can compute the second moment of $t_f$ and correspondingly derived the variance of $t_f$, 
\begin{eqnarray}\label{eq2.8.3.2}
{\rm{Var}} \left(  {t_f} \right)   =\frac{{2D}}{{{\mu ^4}}}\left[ 2D\left(  {{e^{\mu {x_0}/D}} + {e^{2\mu {x_0}/D}} - 2} \right)   \right.  \nonumber \\ \left. - \mu {x_0}\left(  {4{e^{\mu {x_0}/D}} + 1} \right)  \right].
\end{eqnarray}
The variance of $t_f$ diverges as $\mu^{-3}$ in the limit $\mu \to 0$. Also, this  variance is a non-monotonic function of $\mu$ and reaches a minimum at $\mu=1.55094...$ (in units of $D/x_0$). 

\section{exit probability and extreme displace}\label{sec4}
Let us denote by $x_{\max}={\max \left\{ {x(t),0 \leq t \leq {t_f}} \right\} }$ the maximum value of positions of the particle during the first-passage process, the cumulative probability ${\rm{Prob}}\left[ x_{\max} \leq M \right]$ is equivalent to the exit probability through the origin, where the particle diffuses in an interval $\left[0, M \right] $ with absorbing boundaries at both ends.

Consider the Brownian dynamics starting from $x_0 \in \left[  0, M\right] $ and both ends of the interval are absorbing boundaries. Let us denote by $\mathcal{E}(x_0)$ the exit probability that the particle exits the interval for the first time through the origin without hitting the boundary at $x=M$, i.e., the probability that the maximum before the first-passage time is less than or equal to $M$. The backward equation for $\mathcal{E}(x_0)$ reads \cite{redner2001guide}
\begin{eqnarray}\label{eq1.1}
D\frac{{{d^2}\mathcal{E}\left( {{x_0}} \right)}}{{dx_0^2}} -\mu {\rm{sgn}} (x-x_c) \frac{{d \mathcal{E}\left( {{x_0}} \right)}}{{d{x_0}}} = 0,
\end{eqnarray} 
subject to the boundary conditions
\begin{eqnarray}\label{eq1.2}
\mathcal{E}(0)=1, \quad \mathcal{E}(M)=0.
\end{eqnarray} 
Using the boundary conditions in Eq.(\ref{eq1.2}) and continuity conditions at $x=x_c$: $\mathcal{E}(x_c^{+})=\mathcal{E}(x_c^{-})$ and $\mathcal{E}'(x_c^{+})=\mathcal{E}'(x_c^{-})$, Eq.(\ref{eq1.1}) can be solved for $x<x_c$ and $x>x_c$ (assuming $0<x_c<M$), separately, 
\begin{eqnarray}\label{eq1.3}
\mathcal{E}\left( {{x_0}};x_c \right) =\left\{ \begin{array}{lll}\frac{{{e^{2\hat \mu M}} + {e^{ - 2\hat \mu \left( {{x_0} - 2{x_c}} \right)}} - 2{e^{2\hat \mu {x_c}}}}}{{{e^{2\hat \mu M}} - 2{e^{2\hat \mu {x_c}}} + {e^{4\hat \mu {x_c}}}}},    & x_0 \leq x_c,  \\
\frac{{{{\text{e}}^{2\hat \mu M}} - {{\text{e}}^{2\hat \mu {x_0}}}}}{{{e^{2\hat \mu M}} - 2{e^{2\hat \mu {x_c}}} + {e^{4\hat \mu {x_c}}}}},    & x_0 \geq x_c,  \\ 
\end{array}  \right. 
\end{eqnarray}
where $\hat \mu=\mu/(2D)$ again. For the special case when $x_c=x_0$, Eq.(\ref{eq1.3}) reduces to 
\begin{eqnarray}\label{eq1.4}
\mathcal{E}( {{x_0}};x_0 ) =\frac{{\sinh \left[ {\hat \mu ( {M - {x_0}} )} \right]\left[ {\cosh ( {\hat \mu{x_0}} ) - \sinh ( {\hat \mu {x_0}} )} \right]}}{{\sinh ( {\hat \mu M} ) - \cosh ( {\hat \mu M} ) + \cosh \left[ {\hat \mu ( {M - 2{x_0}} )} \right]}}. \nonumber \\
\end{eqnarray}

\begin{figure}
	\centerline{\includegraphics*[width=1.0\columnwidth]{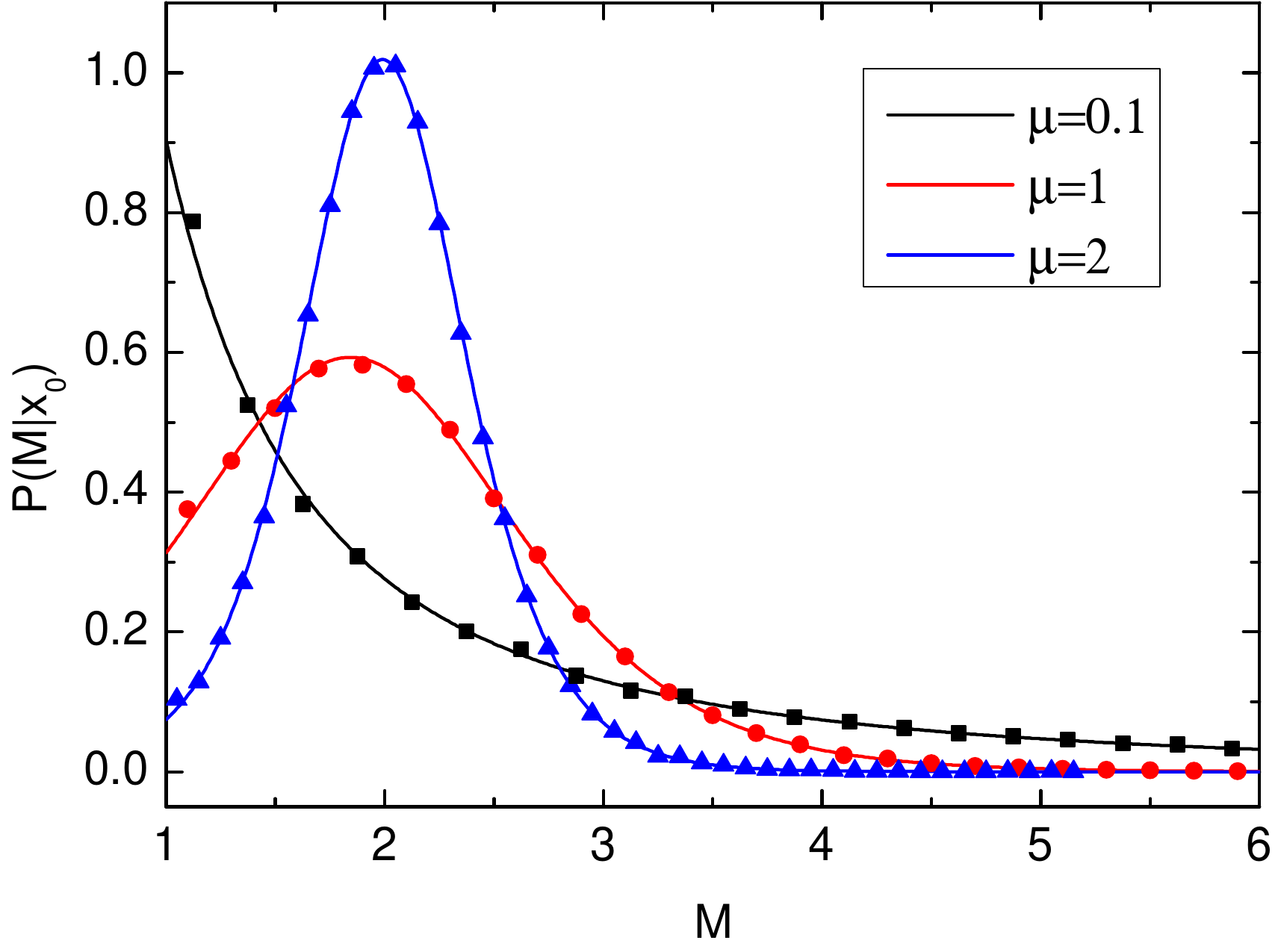}}
	\caption{The marginal distribution $P(M|x_0)$ of the extremal position $M$ of Brownian motion before passing through the zero for three different $\mu$, where $x_0=1$ and $D = 1/2$. The lines and symbols correspond to the theoretical and simulation results, respectively. \label{fig3}}
\end{figure}

Differentiating Eq.(\ref{eq1.4}) with respect to $M$ gives the probability density function of $M$,
\begin{eqnarray}\label{eq1.5}
P( {M|{x_0}} ) = \frac{{\hat \mu \sinh ( \hat \mu {x_0} )  \left[ {\cosh ( {\hat \mu {x_0}} ) - \sinh ( {\hat \mu {x_0}} )} \right]}}{{{{\left( {\sinh ( {\hat \mu M} ) - \cosh ( {\hat \mu M} ) + \cosh \left[ {\hat \mu ( {M - 2{x_0}} )} \right]} \right)}^2}}}.\nonumber \\
\end{eqnarray} 
The marginal distribution of the extreme displacement $M$ is shown as Fig.\ref{fig3} for different $\mu$, where we have considered the case when $x_c=x_0$. In the limit $\mu \to 0$, the potential is absent and the distribution of $M$ is $P(M|x_0)=x_0/M^2$, which decays with $M$ in a power-law way \cite{randon2007distribution}. With the increment of $\mu$, $P(M|x_0)$ can evolves into a nonmonotonic function of $M$, with a peak at an intermediate value of $M$. In the limit $\mu \to \infty$, $P(M|x_0)$ converges to a delta function, showcasing an infinitely sharp peak at $M=2x_0$. To verify the theoretical results, we have performed extensive numerical simulations, as shown by symbols in Fig.\ref{fig3}. In the simulations, we have used a time step $dt=10^{-4}$ and each
set of data is obtained by averaging over $10^5$ first-passage trajectories. Clearly, the simulations agree well with the theory.

\begin{figure}
	\centerline{\includegraphics*[width=1.0\columnwidth]{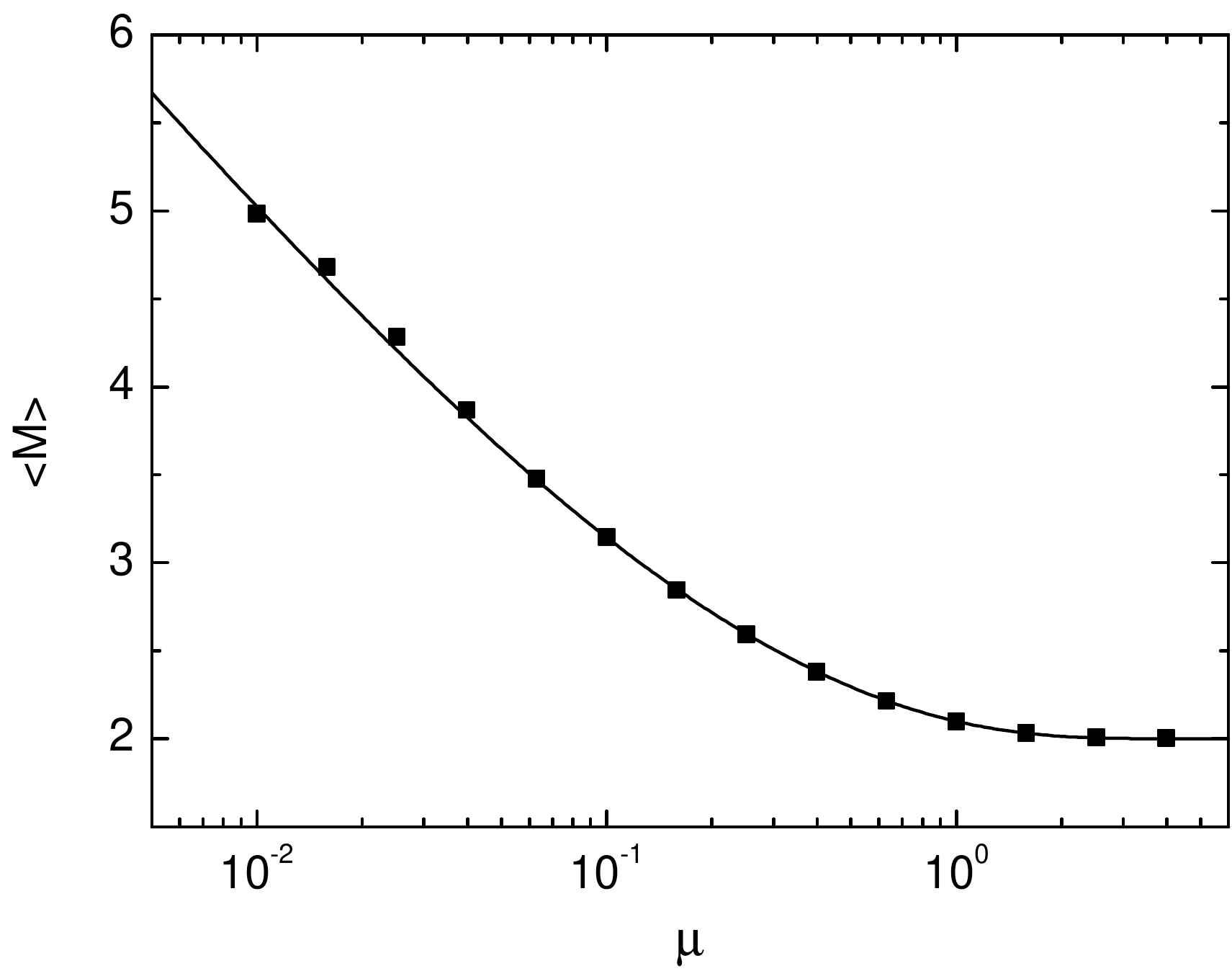}}
	\caption{The expected value $\langle  M \rangle$ of the extreme displacement $M$ as a function of $\mu$, where $x_0=1$ and $D = 1/2$. The lines and symbols correspond to the theoretical and simulation results, respectively. \label{fig2}}
\end{figure}

The expectation of $M$ can be computed as
\begin{eqnarray}\label{eq1.6}
\langle  M \rangle &=& \int_{x_0}^{\infty}P\left( {M|{x_0}} \right) dM  \nonumber \\&= &
{x_0} + \frac{{\log \left[ {\frac{1}{2}{\rm{csch}}\left( {\hat \mu {x_0}} \right)} \right] - \hat \mu {x_0}}}{{\hat \mu \left[ {\coth \left( {\hat \mu{x_0}} \right) - 3} \right]}}.
\end{eqnarray} 
In the limit of $\hat \mu \to 0$, $\langle  M \rangle$ diverges as $\langle  M \rangle \sim -x_0 \ln \left( 2x_0 \hat \mu \right) $. For $\hat \mu > 0$, $\langle  M \rangle$ is finite and decreases monotonically as  $\hat \mu$ increases, and converges to $2 x_0$ in the limit of $\hat \mu \to \infty$. The theoretical and simulation results are shown in Fig.\ref{fig2}, where $x_0=1$ and $D = 1/2$ are fixed. The agreements between them are excellent.

\section{joint distribution of $m$ and $t_m$ in the Laplace domain and the expectation of $t_m$ }\label{sec5}
Let us define $P(M,t_m|x_0)$ as the joint probability density function that the RBM reaches its maximum $M$ at time $t_m$ before passing through the origin for the first time $t_f$, providing that the Brownian starts from the position $x_0$ ($>0$). To compute the joint distribution $P(M,t_m|x_0)$, we can decompose the trajectory into two parts: a left-hand segment for which $0<t<t_m$, and a right-hand segment for which $t_m<t<t_f$, as shown in Fig.\ref{fig0}. Due to the Markovian property of the Brownian dynamics, once the position of the particle is specified at $t_m$, the weights of the left and the right segments become completely independent and the total weight is just proportional to the product of the weights of the two separate segments. For the first segment, we have a process that propagates from $x_0$ at $t=0$ to $M$ at $t=t_m$ without hitting the origin. 
For the second segment, the process propagates from $M$ at $t_m$ to $0$ at $t_f$, where $t_f \geq t_m$ without crossing the level $M$ and the level $0$ in between. The restriction of staying in an interval can be implemented by placing absorbing boundary conditions at $x=0$ and $x=M$. However, the absorbing boundary at $x=M$ will forbid the process to reach $x=M$ at $t_m$. To resovle this difficulty, we introduce a small cut-off $\epsilon$ and define the particle as reaching $M-\epsilon$ at $t_m$. This allows us to consistently apply absorbing boundaries at $x=0$ and $x=M$ over the entire time interval 
$\left[0,t_f\right]$. Finally, we will take the limit $\epsilon \to 0$ in a manner that ensures the physical consistency of the result. Such a limiting cut-off procedure has been used in the literature for similar problems \cite{majumdar2004exact,randon2007distribution,majumdar2008time,PhysRevE.103.022135,guo2023extremal,guo2024extremal,huang2024extreme}.  Below we will employ the same procedure for this problem.

Let us first consider the first segment $0<t<t_m$. The statistical weight of the first segment thus equals to the propagator $G(M-\epsilon,t_m|x_0)$.  For the second segment, its statistical weight is given by the exit probability $\mathcal{E}(M-\epsilon;x_0)$. Therefore, the joint probability density $P(M,t_m|x_0)$ can be written as the product of the statistical weights of two segments \cite{randon2007distribution}, 
\begin{eqnarray}\label{eq4.1}
{P}( {M,{t_m}|{x_0}} ) = \lim_{\epsilon \to 0} \mathcal{N}( {{x_0},\epsilon} ){G}( {M - \epsilon,{t_m}|{x_0}} ){\mathcal{E}}( {M - \epsilon};x_0 ), \nonumber \\
\end{eqnarray}
where the normalization factor $\mathcal{N}( {{x_0},\epsilon} )$  will be determined later. 

One can compute the Laplace transform of $G(M-\epsilon,t_m|x_0)$ in terms of Eq.(\ref{eq2.8}) and ${\mathcal{E}_r}\left( {M - \epsilon};x_0 \right) $ by Eq.(\ref{eq1.3}). In the leading order in $\epsilon$, they are
\begin{eqnarray}\label{eq4.2}
	&&{{\tilde G}}( {M - \epsilon,s|{x_0}} ) \nonumber \\&=& \frac{{{\omega _s}\sinh ( {{\omega _s}{x_0}} )\left( {\sinh \left[ {\hat \mu( {M - {x_0}} )} \right] - \cosh \left[ {\hat \mu( {M - {x_0}} )} \right]} \right)}}{{D\left( {\hat \mu\cosh ( {{\omega _s}M} ) - \hat \mu\cosh \left[ {{\omega _s}( {M - 2{x_0}} )} \right] - {\omega _s}\sinh ( {{\omega _s}M} )} \right)}}\epsilon,\nonumber \\
\end{eqnarray}
and
\begin{eqnarray}\label{eq4.3}
	{\mathcal{E}}\left( {M - \epsilon};x_0 \right) =\frac{{2\hat \mu {e^{2\hat \mu M}}}}{{{e^{2\hat \mu M}} - 2{e^{2\hat \mu {x_0}}} + {e^{4\hat \mu {x_0}}}}}\epsilon.
\end{eqnarray}
	
It is convenient to perform the Laplace transform for $P_r(M,t_m|x_0)$ with respect to $t_m$, 
\begin{eqnarray}\label{eq4.3.1}
	\tilde{P}_r(M,s|x_0)&=&\int_{0}^{\infty} d t_m e^{-s t_m} P_r(M,t_m|x_0) \nonumber \\&=& \mathcal{N}( {{x_0},\epsilon} ){\tilde{G}_r}( {M - \epsilon,{s}|{x_0}} ){\mathcal{E}_r}( {M - \epsilon} ). 
\end{eqnarray}
Letting $s \to 0$, the left-hand side of Eq.(\ref{eq4.3.1}) is just the marginal distribution $P_r(M|x_0)$, which yields
\begin{eqnarray}\label{eq4.4}
	{P_r}( {M|{x_0}} ) &=& \int_0^\infty d{t_m}{P_r}( {M,{t_m}|{x_0}} ) \nonumber \\ &=&  \mathcal{N}( {{x_0},\epsilon} ){{\tilde G}_r}( {M - \epsilon,0|{x_0}} ){\mathcal{E}_r}( {M - \epsilon} ).
\end{eqnarray}
Substituting Eq.(\ref{eq1.5}), Eq.(\ref{eq4.2}) and Eq.(\ref{eq4.3}) into Eq.(\ref{eq4.4}), we obtain
\begin{eqnarray}\label{eq4.5}
	\mathcal{N}( {{x_0},\epsilon} ) =\frac{D}{\epsilon^2},
\end{eqnarray}
which is independent of the starting position $x_0$. In fact, $\mathcal{N}$ can be also obtained by the normalization condition of $P_r(M|x_0)$, i.e., $\int_{x_0}^{\infty} dM P_r(M|x_0)=1$. This can be easily accomplished by integrating the second line of Eq.(\ref{eq4.4}) over $M$. The two methods produce the consistent result.

Inserting  Eq.(\ref{eq4.2}), Eq.(\ref{eq4.3}) and Eq.(\ref{eq4.5}) into Eq.(\ref{eq4.3.1}), we obtain the joint distribution $P(M,t_m|x_0)$ in the Laplace space,
\begin{widetext}
\begin{eqnarray}\label{eq4.5.1}
	\tilde{P}(M,s|x_0)  =\frac{{{\omega _s}\sinh ( {{\omega _s}{x_0}} )\left( {\sinh \left[ {\hat \mu( {M - {x_0}} )} \right] - \cosh \left[ {\hat \mu( {M - {x_0}} )} \right]} \right)}}{{\left( {\hat \mu\cosh ( {{\omega _s}M} ) - \hat \mu\cosh \left[ {{\omega _s}\left( {M - 2{x_0}} \right)} \right] - {\omega _s}\sinh ( {{\omega _s}M} )} \right)}} \frac{{2\hat \mu {e^{2\hat \mu M}}}}{{{e^{2\hat \mu M}} - 2{e^{2\hat \mu {x_0}}} + {e^{4\hat \mu {x_0}}}}}.
\end{eqnarray}
\end{widetext}

\begin{figure}
	\centerline{\includegraphics*[width=1.0\columnwidth]{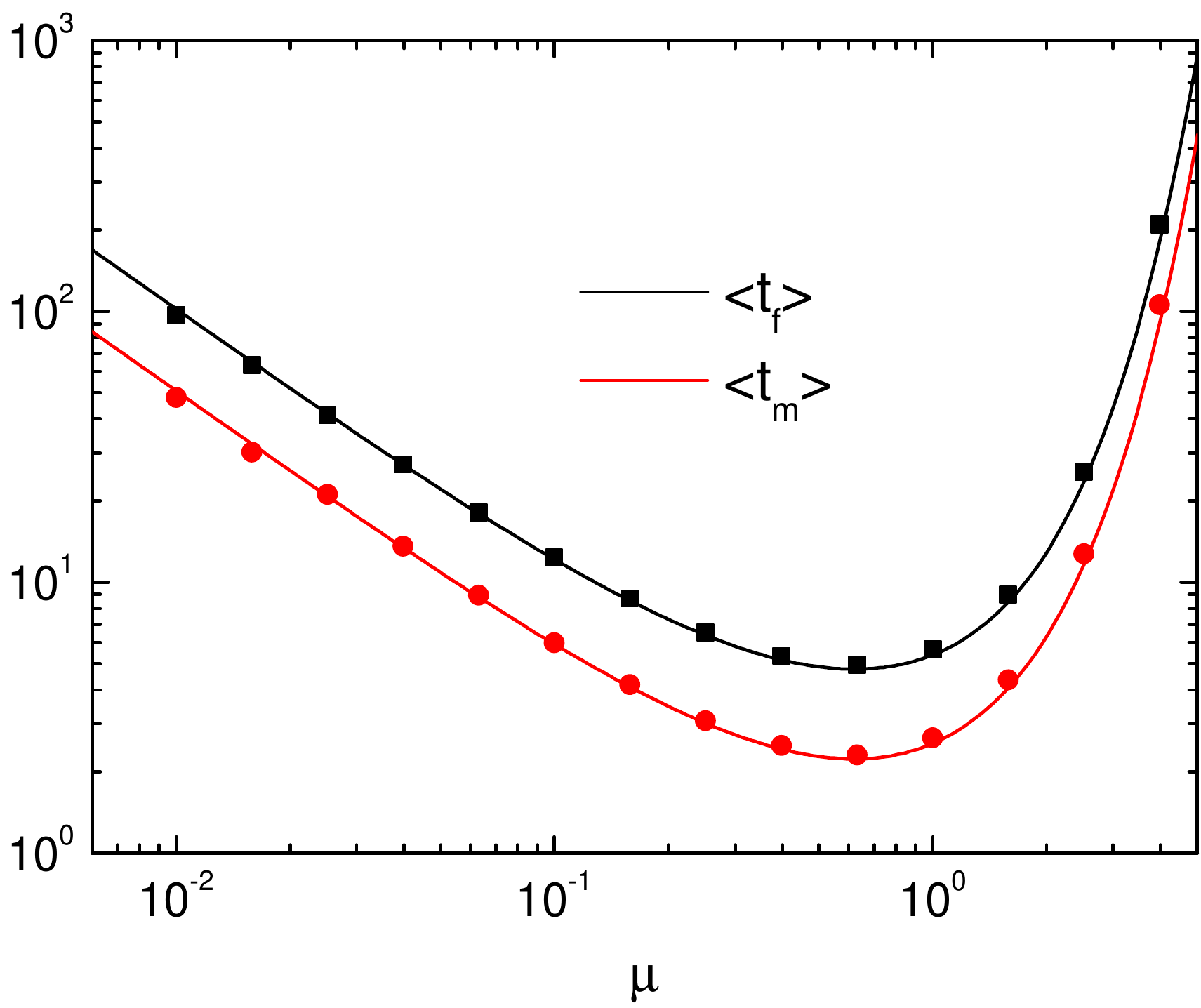}}
	\caption{The mean first-passage time  $\langle t_f \rangle$ and the mean time $\langle t_m \rangle$ of $t_m$ at which the Brownian particle reaches its maximum position before its first passage through the origin, having starting from $x_0=1$, as  functions of $\mu$, where $D = 1/2$. The lines and symbols correspond to the theoretical and simulation results, respectively. \label{fig1}}
\end{figure}

To obtain the joint distribution $P(M,t_m|x_0)$, one has to invert the Laplace transformation for Eq.(\ref{eq4.5.1}) with respect to $s$. Unfortunately, it turns out to be a challenging task. Nevertheless, one may expect to obtain explicitly the statistics of $t_m$, such as the expectation value of $t_m$. To achieve this, by integrating Eq.(\ref{eq4.3.1}) over $M$ from $x_0$ to $\infty$, one obtain the Laplace transform of the marginal distribution $P(t_m|x_0)$,   
\begin{eqnarray}\label{eq4.6}
{{\tilde P}}( {s|{x_0}} ) &=& \int_0^\infty  {d{t_m}} {e^{ - s{t_m}}}{P_r}( {{t_m}|{x_0}} ) \nonumber \\&=& \int_{x_0}^{\infty} dM \tilde{P}_r(M,s|x_0).
\end{eqnarray}
In particular, the expectation  of the time $t_m$ is given by
\begin{eqnarray}\label{eq4.7}
\langle t_m \rangle&=&-\lim_{s \to 0} \frac{\partial \tilde{P}(s|x_0)}{\partial s}\nonumber \\ &=&\frac{{{2 D^2}{{\text{e}}^{\mu {x_0}/D}}}}{{{\mu ^2}}}\left( {1 - \frac{{\mu {x_0}/D}}{{{e^{\mu {x_0}/D}} - 1}}} \right).
\end{eqnarray}
We observe that there exist an optimal $\mu=\mu_*$ minimizing the  $\langle t_m \rangle$. By taking the derivative of $\langle t_m \rangle$ with respect to $\mu$, we find that the transcendental equation satisfied by $\mu_*$, 
\begin{eqnarray}\label{eq4.8}
{y_*^2} + {{{e}}^{y_*}}\left( {{{{e}}^{y_*}} - 1} \right) - 2{\left( {{{{e}}^{y_*}} - 1} \right)^2} = 0, 
\end{eqnarray}
where $y_*=\mu_* x_0/D$. Numerical solution of Eq.(\ref{eq4.8}) shows that 
\begin{eqnarray}\label{eq4.9}
\mu_* \simeq 1.24011 \frac{D}{x_0},
\end{eqnarray}
which is slightly less than the optimal $\mu_{{\rm{opt}}}$ in Eq.(\ref{eq2.8.4}) at which the mean first-passage time is a minimum. As shown in Fig.\ref{fig1}, we plot $\langle t_f \rangle$ and $\langle t_m \rangle$ as functions of $\mu$, where $x_0=1$ and $D = 1/2$ are kept unchanged. The lines and symbols are the theoretical and simulation results, respectively. They are consistent with each other.

\section{conclusions}\label{sec6}
We have shown that how the linear potential affects first-passage properties and the extreme value statistics of the confined Brownian motion during the first-passage processes. The mean first-passage time $\langle {t_f} \rangle $ through the origin nonmonotonically varies with the amplitude $\mu$ of the potential when the particle starts from the position of the potential well. There exists an optimal value of $\mu$, ${\mu_{{\rm{opt}}}} \simeq 1.24468 {D}/{x_0}$, at which the mean first-passage time is a minimum. As $\mu$ approaches zero, $\langle {t_f} \rangle \sim x_0/\mu$, and diverges in the limit of $\mu \to 0$ akin to a free Brownian motion. The distribution $P( { M |{x_0}} )$ of the extremal displacement before passing through the origin has been derived explicitly. The shape of the distribution changes from  monotonically deceasing to nonmonotonic functions of $M$ as $\mu$ is increased. The expectation $\langle M \rangle $ is divergent as $\mu \to 0$, but becomes finite when the confined potential is present. As $\mu$ increases, $\langle M \rangle $ decreases monotonically and converges to $2x_0$ as $\mu \to \infty$. By using the path decomposition technique, we also obtain the distribution in the Laplace space of the time $t_m$ at which the extreme position $M$ is reached, from which we can derive the moments of $t_m$. In particular, we find that the expected value of $t_m$ is also a nonmonotonic function of $\mu$. There is a unique $\mu_* \simeq 1.24011 {D}/{x_0}$ at which $\langle t_m \rangle $ is a minimum.

As an extension for the EVS of Brownian motion under the linear potential, it would be interesting to study the EVS during the first-passage process confined various nonlinear potentials, such as the power-law potentials, not only quadratic potential, or a slower growth logarithmic potential and so on. Due to the growing interest in stochastic resetting \cite{evans2020stochastic}, the impact of resetting on EVS of first-passage trajectories of Brownian motion may be also a compelling direction. Particularly, Brownian motion subject to an intermittent potential-serving as a natural physical realization of stochastic resetting-provides an ideal framework to address this problem \cite{mercado2020intermittent}. In these cases, the EVS for the first-passage trajectories are still open.

\begin{acknowledgments}
This work is supported by the National Natural Science Foundation of China (11875069), the Key Scientific Research Fund of Anhui Provincial Education Department (2023AH050116), and the Educational Department Fund for Distinguished Young Scholars of Anhui Province (No. 2023AH020023)
\end{acknowledgments}


\end{document}